\documentclass[doublecol]{epl2_bioRxiv}

\usepackage{amsmath,bm,amsfonts}
\usepackage{amssymb}

\title{Evolution of populations expanding on curved surfaces}
\shorttitle{Evolution of populations expanding on curved surfaces} 

\author{Daniel A.\ Beller\inst{1,2} \and Kim M. J. Alards\inst{3} \and Francesca Tesser\inst{3,4} \and Ricardo A.\ Mosna\inst{5} \and Federico Toschi\inst{3} \and Wolfram M\"obius\inst{6,7}}
\shortauthor{D.A.\ Beller, \etal}

\institute{                    
  \inst{1} Department of Physics, University of California -- Merced, CA 95343, USA\\
  \inst{2} School of Engineering, Brown University -- Providence, RI 02912, USA\\
  \inst{3} Department of Applied Physics, Eindhoven University of Technology --  P.O. Box 513, 5600 MB Eindhoven, The Netherlands\\
  \inst{4} Laboratoire de Physique et M\'ecanique des Milieux H\'et\'erog\`enes (PMMH), CNRS, ESPCI Paris, PSL Research University, Sorbonne Universit\'e, Univ. Paris Diderot -- Paris, 75005, France\\
  \inst{5} Departamento de Matem\'atica Aplicada, Universidade Estadual de Campinas -- 13083-859, Campinas, SP, Brazil\\
  \inst{6} Living Systems Institute, University of Exeter -- Exeter, EX4 4QD, UK\\
  \inst{7} Physics and Astronomy, College of Engineering, Mathematics and Physical Sciences, University of Exeter -- Exeter, EX4 4QL, UK
}

\abstract{
The expansion of a population into new habitat is a transient process that leaves its footprints in the genetic composition of the expanding population.
How the structure of the environment shapes the population front and the evolutionary dynamics during such a range expansion is little understood. Here, we investigate the evolutionary dynamics of populations consisting of many selectively neutral genotypes expanding on curved surfaces. Using a combination of individual-based off-lattice simulations, geometrical arguments, and lattice-based stepping-stone simulations, we characterise the effect of individual bumps on an otherwise flat surface. Compared to the case of a range expansion on a flat surface, we observe a transient relative increase, followed by a decrease, in neutral genetic diversity at the population front. In addition, we find that individuals at the sides of the bump have a dramatically increased expected number of descendants, while their neighbours closer to the bump's centre are far less lucky. Both observations can be explained using an analytical description of straight paths (geodesics) on the curved surface. Complementing previous studies of heterogeneous flat environments, the findings here build our understanding of how complex environments shape the evolutionary dynamics of expanding populations.
}

\begin{document}

\maketitle
The expansion of a population into new habitat links population growth with spatial structure. From a demographic perspective, such a range expansion is a transient event whose dynamics is governed by population growth at the frontier and thus is closely linked to studies of front propagation \cite{vanSaarloosPR2003}. From an evolutionary perspective, a range expansion has lasting consequences, with the evolutionary dynamics at the front being imprinted into the population at long times \cite{ExoffierTrendEcoEvo2008}. Many works have studied this effect over the last 15 years with the key finding that a range expansion leads to a small effective population size at the front and thus amplifies the effects of randomness, leading to a decline in genetic diversity and, possibly, the formation of monoclonal regions (see, e.g., Ref.~\cite{EdmondsPNAS2004,KlopfsteinMBE2006,HallatschekTPB2008,HallatschekPNAS2007,ExoffierTrendEcoEvo2008}).

The majority of these studies focuses on homogeneous environments. The effects of environmental heterogeneity on the evolutionary dynamics in two dimensions is only beginning to be understood, mostly in simulation approaches. 
The environments considered range from very regular and simplistic -- useful for illuminating general effects, e.g., Ref.\ \cite{MoebiusPCB2015,BurtonHeredity2008,MonaHeredity2014} -- to real-world patterns that allow description of past or future evolutionary dynamics facilitated by dedicated simulation packages \cite{RayBioinformatics2010,BocediMEE2014}.

In both types of studies, the role of topography on the evolutionary dynamics has so far attracted less attention. We thus know very little about the effect of a surface's shape on the evolutionary dynamics of a range expansion. Recent work examining radially symmetric populations expanding in a time-dependent manner could be generalised to understand spread on radially symmetric surfaces \cite{LavrentovichTPB2015}. More general scenarios have not been considered explicitly, to our knowledge.

Here, we consider an initially linear population front, with a large number of selectively neutral genotypes, encountering an isolated topographical feature. Starting with individual-based off-lattice simulations, we show how geodesics on the curved surfaces affect the structure of the population through the phenomenon of \textit{topographic lensing}. We then introduce a lattice-based simulation approach to quantitatively study the consequences of this lensing on the evolutionary dynamics. 

\section{Results}
\subsection{Individual-based simulations}
\begin{figure}
\onefigure[width=88mm]{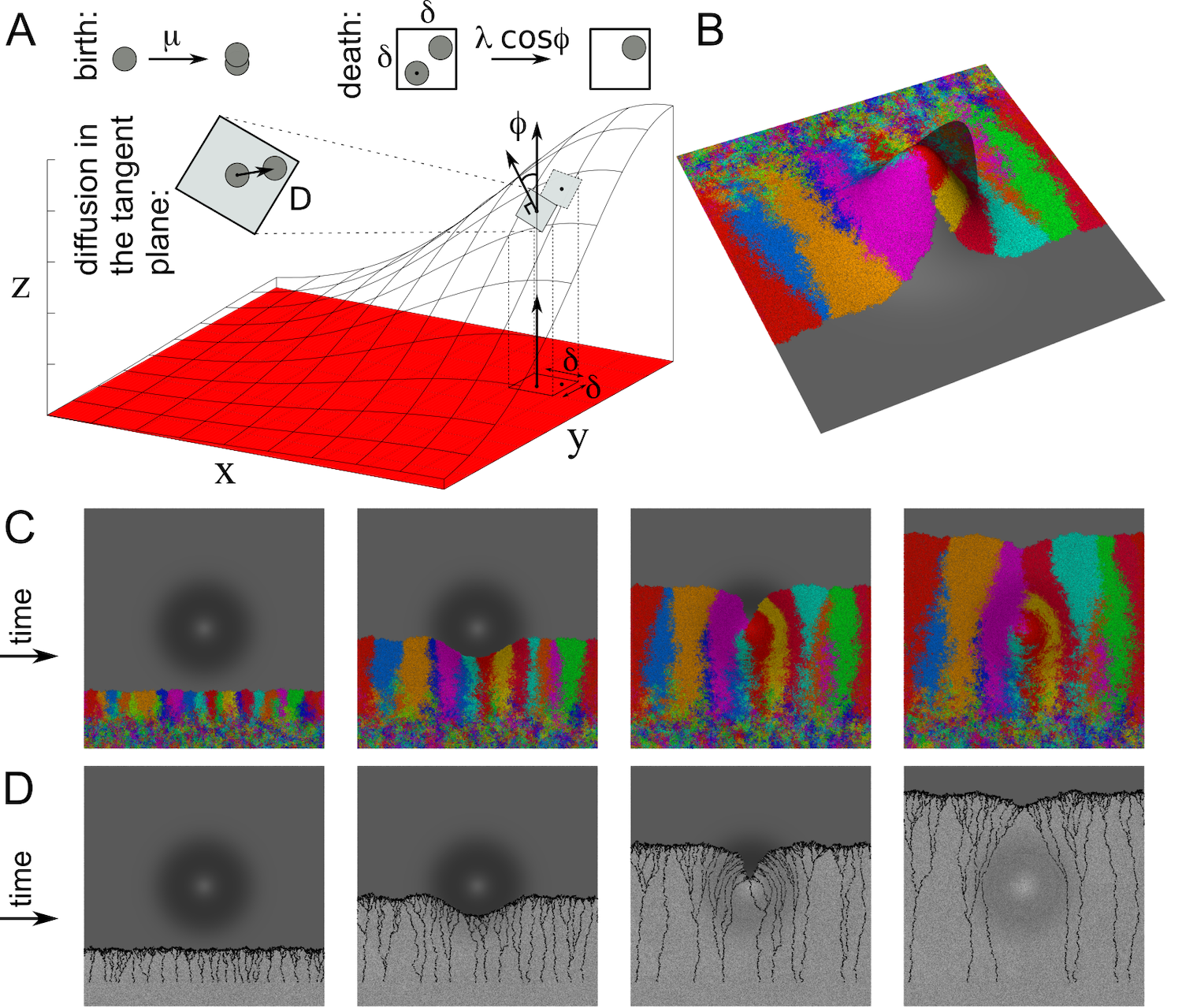}
\caption{\textbf{Simulating population expansions on Gaussian-shaped bumps.} (A) Schematic of the individual-based simulation on a curved surface, with birth rate $\mu$, a death process simulating competition between particles within an interaction area of $\delta^2$ on the projected $xy$ plane, and diffusion coefficient $D$. 
(B) Snapshot of a simulation starting from a linear initial population front, in which each individual carries a distinct genotype marked by a distinct colour.
(C,D) Top view of the same simulation as in panel (B) at equal time intervals, showing (C) genotypes and (D) black dots marking the birth locations
of ancestors of the individuals currently at the front. Collectively these dots trace out the lineages. See Appendix for simulation details and Videos 1,2 for all frames and side views.}
\label{fig1}
\end{figure}

To explore the effects of curved surfaces on a population expansion, we use an off-lattice simulation of individuals undergoing spatial diffusion and birth-and-death processes, following the work by Pigolotti \textit{et al.} \cite{PigolottiTPB2013}, generalized to a surface embedded in three dimensions described by a height function $z=f(x,y)$ (Fig.\,\ref{fig1}A). 


Individuals are duplicated at a birth rate $\mu$. 
A corresponding death rate, simulating competition within an interaction range $\delta$, is given by the number $n$ of other individuals within a square with edge length $\delta$ on the projected $xy$ plane, multiplied by a factor $\lambda\cdot\cos\phi$, where $\phi$ is the angle between the vertical direction and the curved surface's local unit normal vector at the location of the disappearing particle (Fig.\,\ref{fig1}A). The factor $\cos\phi$ occurs because the corresponding region on the surface has a larger area,  $\delta^2 / \cos\phi$. Since the birth event is a purely local process, this rescaling of area does not affect the birth rate.

In addition, particles diffuse on the surface. Diffusion during a time step $dt$ is implemented as two steps of size $\sqrt{2Ddt} \eta_i(t)$, $i=1,2$, along two orthogonal directions $\vec{n}_1$, $\vec{n}_2$ spanning the tangent plane, where the $\eta_i(t)$ are independent white noise factors and $D$ is the diffusion coefficient.  
The particle's new $z$ position is then adjusted so that it lies on the surface, $z=f(x,y)$. 
Note that this simulation scheme for curved topographies requires $f(x,y)$ to vary sufficiently slowly, a condition that is not satisfied everywhere. We will comment on this aspect below.

\subsection{Kinks in fronts and lensing of lineages}
We first simulate an expanding population, starting from an initially populated area with a linear front, encountering a Gaussian-shaped bump; see Fig.\,\ref{fig1}B,C and Video~1. 
While the propagating front approximately maintains its linear shape for some time, upon encountering the bump the front appears perturbed as seen in projection. After the front has passed the top of the bump, a kink forms and then gradually heals. This behaviour is reminiscent of the front perturbations induced by obstacles, where a similar formation and healing of a kink in the front could be explained using a constant-speed propagation argument \cite{MoebiusPCB2015}. 
Below, we will test whether an analogous argument can describe front dynamics in the case of non-flat surfaces.

We give each individual in the initial population a distinct ``genotype'', represented as colours in Fig.\,\ref{fig1}B,C, and inherited without mutation by all descendants.
As the front advances, 
single-colour sectors form which coarsen with time, due to  the small number of individuals contributing to the advancing front \cite{HallatschekPNAS2007,ExoffierTrendEcoEvo2008}. While boundaries between sectors generically display lateral fluctuations, in the absence of selection their average direction is, by symmetry, that of the population front's movement, as long as the sector remains present in the front. On the curved surface, a kink that develops in the front traces out a curve where sectors are rapidly removed from the front. The instance depicted in Fig.\,\ref{fig1}C shows that sectors traversing the bump are squeezed out at the kink, while sectors on the flank are broadened. 
For topographic bumps, as for obstacles  \cite{MoebiusPCB2015}, the fates of genotypes in a front depend on their locations relative to the heterogeneity, a scenario we call ``geometry-enhanced genetic drift''. 

In addition to the genotype, we also track the sequence of birth events connecting each individual at the front to its ancestor in the initial population. The locations of these birth events collectively trace out coalescing lineage trees with roots at the locations of all original ancestors still represented in the front by surviving descendants. The study of population genetics in this reverse-time view is known as coalescent theory \cite{kingman1982coalescent}, and as the `structured coalescent' for spatially structured populations \cite{wilkinson1998genealogy}. Fig.\,\ref{fig1}D and Video~2 display these lineage trees for a simulated expansion over the Gaussian-shaped bump. 

Like the sector boundaries, these lineage curves fluctuate transversely about the front's propagation direction. We will show below that the strong bending of lineages passing over the bump toward the bump's centre, as viewed in projection from above (Fig.\,\ref{fig1}D, third and fourth panels), is a consequence of the topographic curvature. 

\begin{figure}
\onefigure[width=88mm]{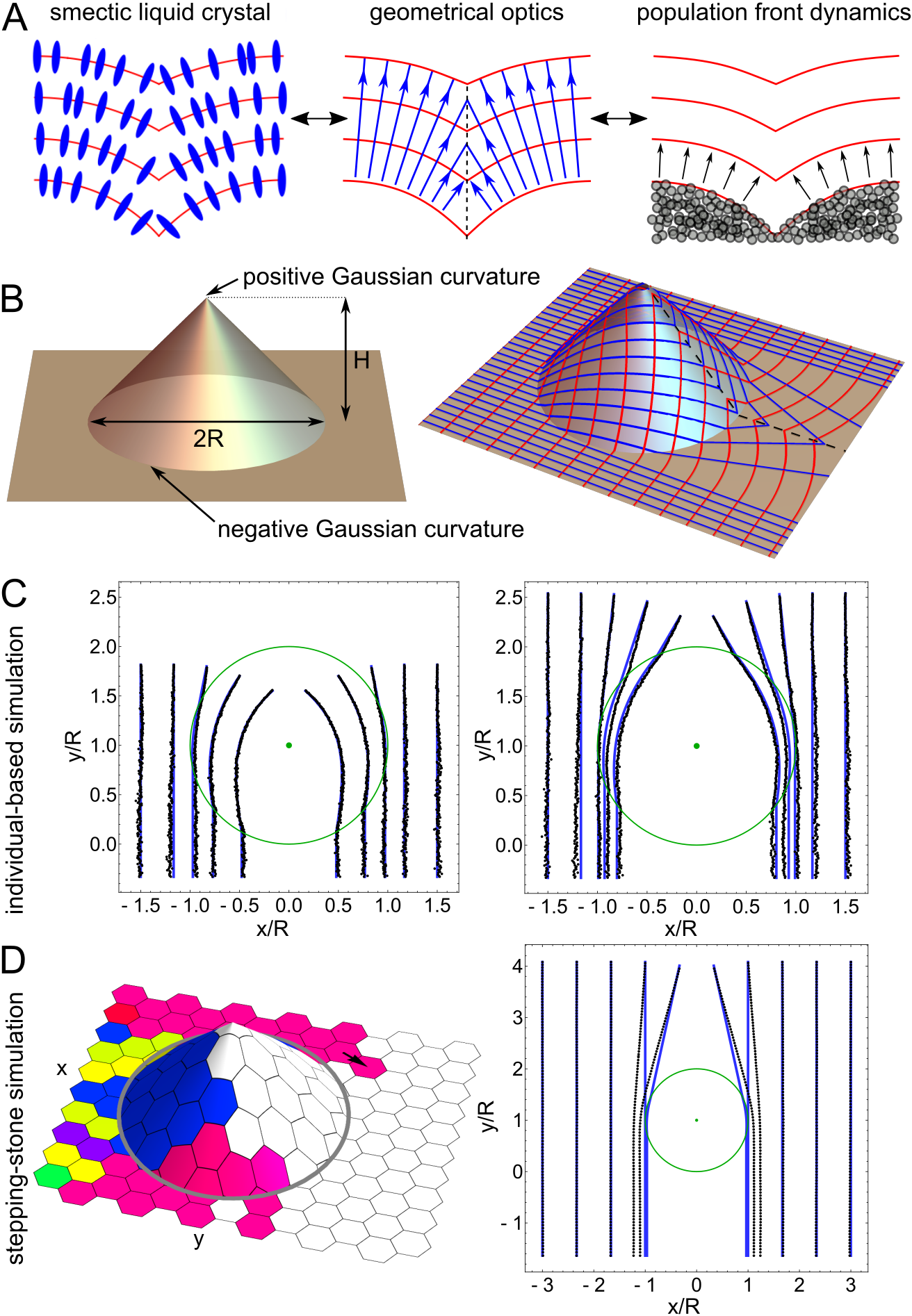}
\caption{\textbf{Equal-time fronts and geodesics on conical bumps.} (A) Schematic illustrations of smectic liquid crystals (left), geometrical optics (centre), and the propagating front of a range expansion (right). (B) Left: Schematic of a conical bump indicating radius $R$ and height $H$ as well as locations of non-vanishing Gaussian curvature. Right: Conical bump with $H=R$, with geodesics (blue), fronts (red), and caustic (black dashed line), following Ref.~\cite{MosnaPRE2012}. (C) Ensemble averages of lineages from individual-based simulations (black dots) and geodesics (blue lines) for two positions of the front for a conical bump with $H=R=1$. (D) Illustration of the stepping-stone simulation (the arrow indicates a ``birth event'' where a marker is copied onto a neighbouring lattice site) as well as a comparison of ensemble-averaged lineages derived from these simulations (black dots) and geodesics (blue lines) for a conical bump with $H=R$.}
\label{fig2}
\end{figure}

\subsection{Geodesics and equal-time fronts on conical bumps}
We are interested in the population's expansion over a bump rising out of a planar surface, with constant propagation speed as measured locally in the curved surface. This propagation is amenable to analytical solution for certain special bump geometries. One such geometry is the cone, for which this problem was studied previously \cite{MosnaPRE2012} in the context of smectic liquid crystals, materials in which rod-like particles organize into equally spaced layers, which may be curved with a finite energetic cost (Fig.\,\ref{fig2}A, left panel) \cite{KlemanBook2002}. 
With the spacing from one smectic layer to the next reinterpreted as propagation during a fixed time interval, the ideal smectic layers are  the same as wavefronts of constant speed propagation within the surface, in the idealization of geometrical optics (Fig.\,\ref{fig2}A, middle panel). The smectic layer normals' integral curves, like the imagined light rays, are geodesics, curves that provide shortest paths within the surface. For the range expansion, if we assume a constant speed of propagation along the population front's local normal direction in the surface, then at equal time intervals the fronts are constructed in exactly the same way as the geometrical optics wavefronts and the idealized smectic layers, with the local front normal direction following geodesics (Fig.\,\ref{fig2}A, right panel).

In understanding how curved topographies affect these geodesics, 
we will see that a central role is played by the surface's Gaussian curvature, $K(x,y) = [f_{xx} f_{yy} - f_{xy}^2]/[(1+f_x^2+f_y^2)^2]$,
where again the surface is given by the height function $z=f(x,y)$, and the subscripts denote spatial derivatives. The importance of the Gaussian curvature can be seen in its effect on the change in spacing $\xi$ between (infinitesimally) nearby geodesics as they propagate forward with time $t$:
\begin{align}
d^2 \xi / dt^2 &= - K\xi \label{eq: gde}\,.
\end{align}
Eq.~\ref{eq: gde} is the \textit{geodesic deviation equation} expressed for two-dimensional manifolds \cite{DoCarmo2016Book,KamienPRE2009}. It shows that geodesics encountering positive Gaussian curvature will be squeezed together, whereas geodesics encountering negative Gaussian curvature will spread apart. Because of the second derivative, this squeezing or spreading continues beyond the region of non-zero Gaussian curvature. A bump thus behaves like a lens for geodesics \cite{KamienPRE2009} (Figs.\ \ref{fig1}B-D, \ref{fig2}B-D), and the effect is termed ``topographic lensing''. 

Using conical bumps as prototypical curved surfaces makes the problem analytically tractable by  concentrating all of the Gaussian curvature at the cone apex ($K>0$) and edge ($K<0$), as depicted in the left panel of Fig.~\ref{fig2}B. 
The geometrical optics front profiles at various times, starting with a flat front on the left, are indicated by red lines in the right panel of Fig.~\ref{fig2}B, with the corresponding geodesics shown in blue, following Ref.\ \cite{MosnaPRE2012}. 
When geodesics entering the cone encounter the cone edge, which is a source of negative Gaussian curvature, they begin to spread apart according to Eq.~\ref{eq: gde}. The rate of spreading is larger the further away the geodesics are from the symmetry axis. When geodesics \textit{exiting} the cone encounter the edge a second time, spreading is accelerated.  Geodesics that never enter the cone remain straight lines in the plane. 

How can geodesics that encounter the cone be continually spreading apart from one another, while they are bounded on either side by straight-line geodesics continuing unperturbed? The answer is that the spreading geodesics end, one after another, on a caustic extending from the cone apex down the cone and onto the plane in the direction of propagation (black dashed line in Fig.\,\ref{fig2}B). The caustic is the trace of the kink in the propagating front. Because the front normal is discontinuous there, the geodesics collide with their mirror-image geodesics from the other side of the cone. Unlike light waves, the population fronts (like smectic layers) cannot have one portion passing through another. Instead, a geodesic encountering the caustic simply ends, and an element of the front is lost to the caustic. All geodesics that enter the cone are doomed to eventually end on the caustic.

\subsection{Ensemble-averaged lineages on a conical bump}
The geodesics provide a deterministic prediction for the front normal directions, about which lineages and sector boundaries are expected to fluctuate, as argued above. While each instance of a simulation as depicted in Fig.~\ref{fig1}B,C is a stochastic process, 
we can test whether there is predictable deterministic behaviour that can be described by the geometrical optics approach illustrated in Figs.~\ref{fig2}A,B. We argue that the deterministic behaviour is revealed by the ensemble average of lineages (see Appendix).



We compute ensemble-averaged lineage trajectories as follows: For all the lineage realizations ending at a given position on the final-time front, we calculate the ``centre of mass'' of those lineages' constituent points within a given time bin. 
On the curved surface, the centre of mass calculation generalises the arithmetic mean, seeking the point on the surface that minimises the sum of squared in-surface distances to the set of points (see Appendix). 

As Fig.~\ref{fig2}C illustrates, the ensemble-averaged lineages terminating at different front positions are indeed well-described by the analytically calculated geodesics. This  heuristically justifies our application of a simulation scheme designed for slowly changing surfaces to the conical bump, whose surface normal is discontinuous at the cone edge and apex. 
We will see below how this average propagation along geodesics is key to understanding the effects of the bump on genetic diversity.

\subsection{Stepping-stone simulations}
To explore the evolutionary consequences of  traversing the bump for a range of parameters, we introduce a coarse-grained, lattice-based simulation scheme that is far more computationally efficient than the individual-based approach. Following a series of work investigating the evolutionary consequences of range expansions on flat surfaces (e.g., \cite{KorolevRevModPhys2010}) we describe the spread of the population by a variant of the stepping-stone model \cite{kimura1964stepping} with one organism per deme (site). Individual sites on a hexagonal lattice can be unoccupied (population not present) or occupied and associated with a genotype. A site does not change its state after becoming occupied. The front propagates as lattice sites copy their genotype markers onto empty neighbours (``daughters'').

To incorporate the topographic bump, we use a hexagonal lattice in the projected horizontal plane with weighted links $w_{i\rightarrow j}$ between all occupied lattice sites at the front and each of their unoccupied neighbours (Fig.~\ref{fig2}D). Here, $w_{i\rightarrow j}$ is the probability that the next birth event will copy the marker of filled site $i$ into a neighbouring empty site~$j$. We choose $w_{i\rightarrow j}\propto 1/(d_{ij} q_{i})$, where $d_{ij}$ is the distance between the two sites as measured on the surface, and $q_i$ is the number of currently empty neighbours of site $i$. Normalisation is realised by considering all possible mother-daughter pairs. On the plane, where $d_{ij}$ is uniform, this procedure reduces to the Eden or ``rough-front'' growth model \cite{Eden1961,JullienPRL1985}
. On a general surface, the weighting inversely with distance $d_{ij}$ gives us, approximately, a uniform speed of front propagation as measured in the surface, allowing us to explore the effects of topographic lensing. Alternative strategies include triangulating the curved surface; see Ref.\ \cite{wang1997biological} for one off-lattice method.

We tracked ancestral lineages as we did for the individual-based simulation approach. Fig.~\ref{fig2}D shows the ensemble-averaged lineages for a cone with $H/R=1$ after the front has passed the cone. While the lensing is visible qualitatively, there is some disagreement between predicted ensemble-averaged lineages and those obtained from the stepping-stone simulation. We attribute this discrepancy to lattice effects. For this reason, care should be taken in making quantitative comparisons between results of these stepping-stone simulations and other approaches.


\begin{figure}
\onefigure[width=88mm]{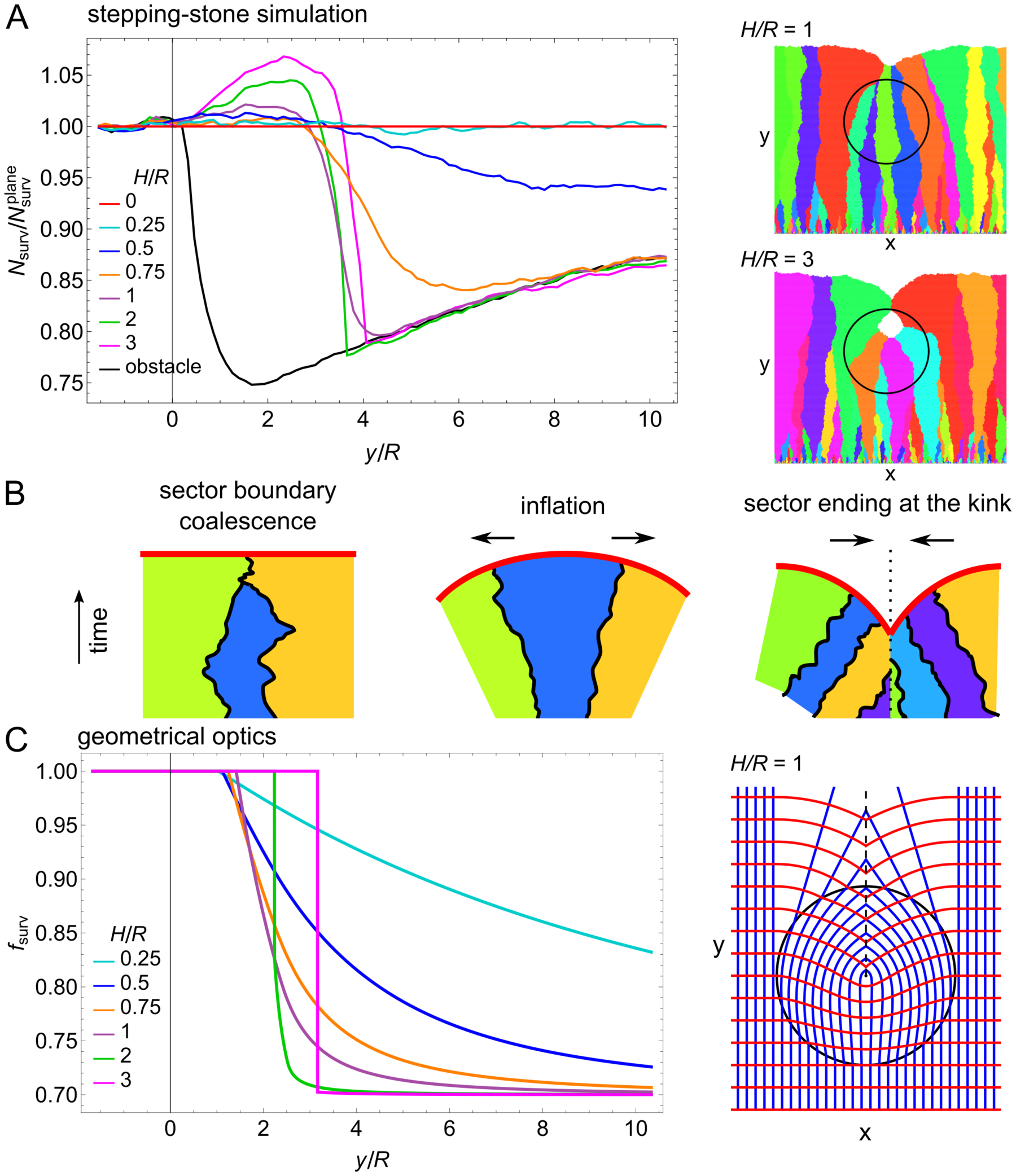}
\caption{\textbf{Accelerated loss of genetic diversity on a conical bump.} (A) Average number of surviving sectors $N_{\mathrm{surv}}$ in the front, as a function of distance $y$ traveled by the front since first encountering the cone edge, normalised by the corresponding quantity on a planar surface, $N_{\mathrm{surv}}^{\mathrm{plane}}$.  Different ratios of cone height $H$ to cone radius $R$ are considered. Images on the right are snapshots from simulations with cones of $H/R=1,3$. (B) Sketches of the two mechanisms for loss of sectors: (left) coalescence of sector boundaries and (right) front coming to a halt through encountering another part of the front. Centre: a cartoon of how inflation counteracts sector loss. (C) Fraction of geodesics $f_{\mathrm{surv}}$ surviving (i.e., not ending in the caustic) as a function of front distance travelled. $H,\,R,\,y$ are defined as in (A).} 
\label{fig3}
\end{figure}

\subsection{Effect of cone on genetic diversity}
Fig.~\ref{fig1}C illustrates loss of genetic diversity, the reduction in number of sectors (i.e., genotypes), due to the bump. 
This effect can be quantified in the stepping-stone model by considering the mean number of sectors surviving at the front as a function of how far the front has progressed, $N_{\mathrm{surv}}(y)$. Even in the absence of curved topography, the number of sectors decreases rapidly with time due to sector coarsening \cite{HallatschekEvolution2010}. To see clearly the effects of the curved surface, we normalize $N_{\mathrm{surv}}(y)$ by the mean number of surviving sectors computed for a linear front in a flat environment that has travelled the same distance, $N_{\mathrm{surv}}^{\mathrm{plane}}(y)$. 
This normalisation gives the relative number of surviving sectors on the surface (Fig.~\ref{fig3}A). As expected, for small ratios $H/R$ of cone height to cone radius, the number of sectors surviving is indistinguishable from that for a flat environment. For taller cones (larger $H/R$), the relative number of surviving sectors first increases, then decreases, and finally $N_{\mathrm{surv}}/N_{\mathrm{surv}}^{\mathrm{plane}}$ appears to converge to a common curve. 
This common curve is the same as that for a disk-shaped obstacle (sites inaccessible to birth events) with the same centre and radius as the cone (black curve in Fig.\,\ref{fig3}A; see also Ref.~\cite{MoebiusPCB2015}). Because sectors entering a tall cone have a vanishing chance of contributing to the front after it passes the cone, the relative number of surviving sectors eventually agrees with the obstacle case, where the same sectors are simply blocked from advancing. The gradual increase of this common curve at late times reflects the diminishing effect of the bump or obstacle as it recedes into the past.  

The decrease in sector number can result from two sector boundaries merging and leaving a sector behind (Fig.~\ref{fig3}B left panel). This is the mechanism for sector coarsening in a flat environment \cite{HallatschekEvolution2010,WeinsteinPCB2017}. Inflation of the front can decrease the rate at which sector boundaries stochastically encounter one another, by deterministically driving sector boundaries apart (Fig.~\ref{fig3}B centre panel), for example in radially growing populations \cite{HallatschekPNAS2007,KorolevRevModPhys2010,lavrentovich2013radial}. On the conical bump, a local inflationary effect occurs when the front encounters the cone edge and geodesics spread apart due to the negative Gaussian curvature.  In consequence, fewer sectors disappear compared to the flat case, causing the (temporary) increase in $N_{\mathrm{surv}}/N_{\mathrm{surv}}^{\mathrm{plane}}$ seen in Fig.~\ref{fig3}A. The obstacle has no such inflationary increase because the portion of the front encountering the obstacle is immediately halted, whereas even for the tallest cones the front continues to advance after encountering the cone edge. While this inflationary effect is evident even for small cones, it is magnified substantially by larger cone height.

What then leads to the subsequent pronounced decrease in the relative number of surviving sectors?
The answer comes from a second mechanism for the loss of sectors: A portion of the population front may come to a halt through self-intersection and cease to contribute to the range expansion (Fig.~\ref{fig3}B right panel). This situation occurs at the kink and corresponds, in the geometrical optics picture, to geodesics ending in the caustic. 
Fig.~\ref{fig3}C displays the fraction of surviving geodesics as a function of distance $y$ that the front has travelled since first encountering the cone. The loss of geodesics to the caustic begins at different $y$ values for different cone heights because the kink  forms later for taller cones. As expected, the loss of geodesics occurs more gradually for shorter cones, but eventually all geodesics that encounter the cone will be lost at very large $y$. Note that the limiting value $f_\textrm{surv}\rightarrow 0.70$ at large $y$ in Fig.~\ref{fig3}C is a consequence of our choice of cone diameter $2R=0.30 L$, where $L$ is the system width (size of initial population). Likewise, the magnitude of the changes in $N_{\mathrm{surv}}/N_{\mathrm{surv}}^{\mathrm{plane}}$ seen in Fig.~\ref{fig3}A depends on the same choice of parameters in the stepping-stone simulation.

Comparing the computed relative number of surviving sectors (Fig.~\ref{fig3}A) to the analytically calculated fraction of surviving geodesics (Fig.~\ref{fig3}C), we observe a qualitative agreement in the onset and strength of the increased rate of loss of genetic diversity for various cone heights. A quantitative comparison is not appropriate because the simple calculation used to make Fig.~\ref{fig3}C does not capture inflationary effects or the role of fluctuations. Future work will seek analytical approaches incorporating these effects.

For tall cones with $H/R>\sqrt{3}$,
some geodesics turn through an angle greater than $\pi$, which sends them looping around the cone apex, as explored in Ref.~\cite{MosnaPRE2012}. 
The implication for range expansions is that the front breaks up into two disconnected portions, one ringing the apex and advancing up the cone, the other progressing down and past the cone as discussed. The former portion of the front will eventually run out of new territory for expansion, and will shrink to zero length. This prediction is confirmed in the stepping-stone simulations for tall cones, where the front breaks up into two disconnected portions; see Fig.~\ref{fig3}A bottom right panel. (Note that we define the front as the set of occupied sites adjacent to one or more empty sites.)  
Eventually the internal front disappears. This peculiarity therefore does not play a role at long times. 

\begin{figure}
\onefigure[width=78mm]{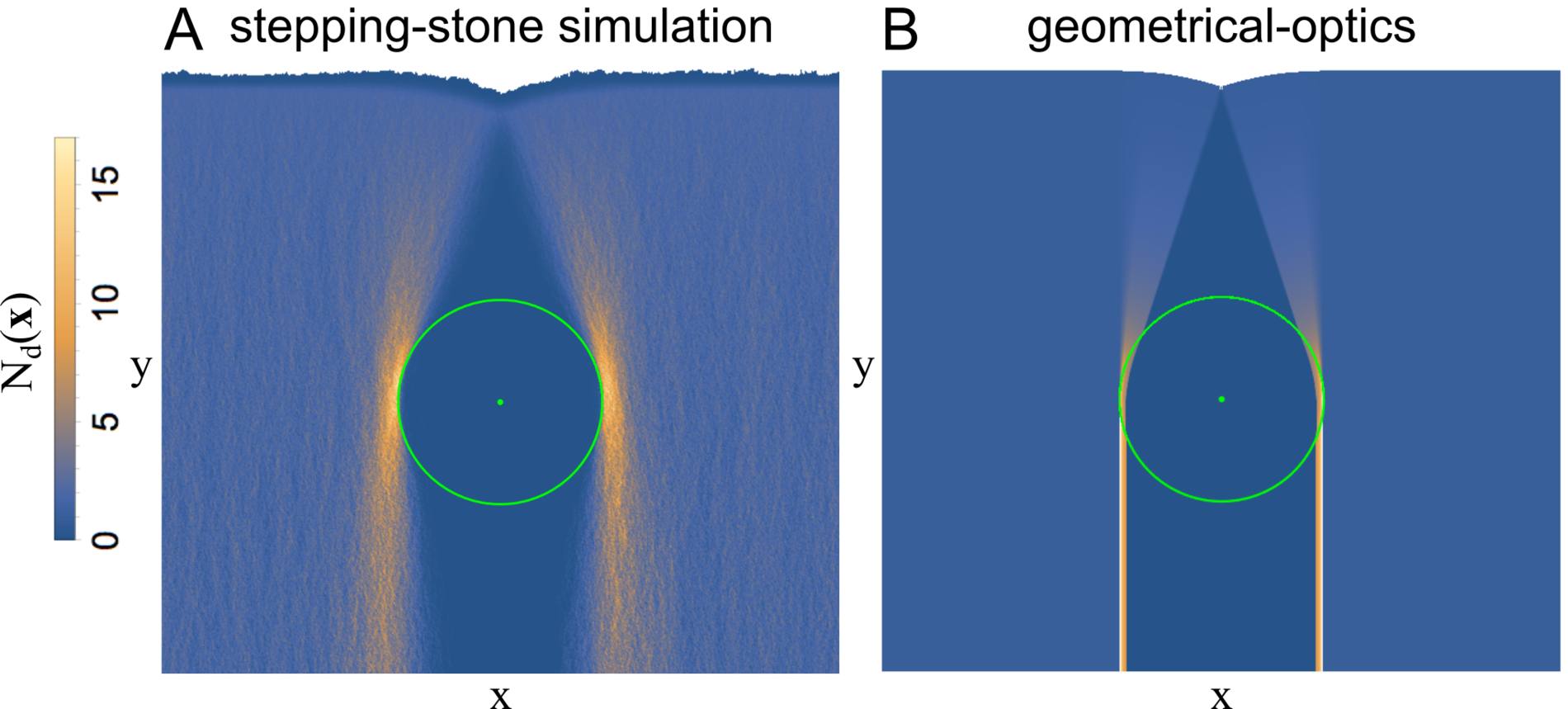}
\caption{\textbf{Location-dependent contribution to genetic make-up of the front.} The expected number of living descendants $N_d(\mathbf{x})$ as a function of location after the front has engulfed the cone, as determined (A) from stepping-stone simulations and (B) using the geodesic deviation equation, Eq.\,\ref{eq: gde}. The cone edge and apex locations are indicated in green.}
\label{fig4}
\end{figure}

\subsection{Location-dependent contribution to genetic make-up of the front}
With fewer sectors surviving, the average sector size must be larger after the front has passed the bump. Fig.~\ref{fig1}B suggests that this increase in average size is not shared equally by all sectors, but rather that some particular sectors benefit substantially from the topography. At a given time, an individual's expected number of ``living descendants'' in the front, $N_d(\mathbf{x})$, depends strongly on that individual's location relative to the cone. We use the stepping-stone simulation approach to quantify, with $N_d(\mathbf{x})$, how topography affects which regions contribute most to the genetic make-up of the front at later times. 

Figure~\ref{fig4}A shows the computed  $N_d(\mathbf{x})$ for a cone with $H/R=1$. Lattice sites close to the sides of the cone edge (and their ancestors below) make a substantially enhanced contribution to the front after it has passed the bump. Qualitatively, this can be understood as follows: Individuals whose descendants at the front follow diverging geodesics have a larger number of descendants at the front at later times. This effect is most pronounced for those geodesics encountering the sides of the cone edge since the effect of divergence is largest for those. However, having a larger number of descendants at the front is contingent on the individuals actually contributing to the front, i.e., that geodesics have not ended in the caustic yet. For long times, this is only the case for sites close to the sides of the cone. Thus, these sites have the largest expected number of descendants at long times. Their neighbours closer to the cone centre have dramatically fewer descendants at the front because geometry has instead pushed their descendants into the caustic.
The above argument can be made quantitative using the geodesic deviation equation (Eq.~\ref{eq: gde}). The result is illustrated Fig.~\ref{fig4}B and predicts the salient features of the simulation results shown in Fig.~\ref{fig4}A. 


\subsection{Toward the genetic consequences of rough topographies}
\begin{figure}
\onefigure[width=80mm]{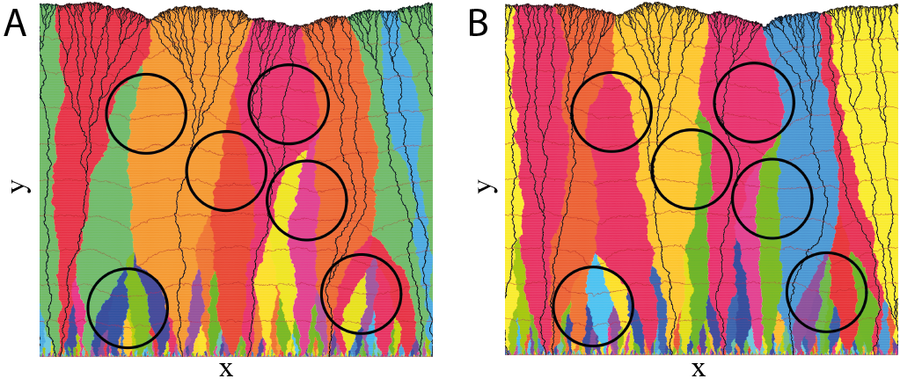}
\caption{\textbf{Effect of multiple conical bumps on genetic composition of an expanding population.} Visualisations of two stepping-stone simulations on a surface with six conical bumps (outlined by black circles). Colours represent genotypes, black curves represent ancestral lineages, and brown curves depict the position of the front at various (earlier) times.}
\label{fig5}
\end{figure}

So far, we have limited ourselves to isolated perturbations of an otherwise flat environment. While an extension to complex, rough environments is beyond the scope of this letter, we briefly analyse simulations with one surface containing a few bumps. Fig.~\ref{fig5} shows the lineages together with the sectors for two instances of the simulation. For bumps close to the initial front (bottommost circles in Fig.~\ref{fig5}) we observe the rapid loss of sectors in caustics as discussed above. For bumps passed by the front at later times  this effect is less prominent, because sectors have coarsened enough that only a few sectors encounter each bump. Most lineages that survive for long times pass around the edges of, rather than over, the conical bumps. This tendency is strongest for bumps encountered longest ago in the past. Through multiple iterations of the effect illustrated in Fig.\,\ref{fig4}, the placement of multiple bumps can therefore favour some tortuous lineage trajectories diverted through the spaces between them. 

\section{Discussion}

We have applied an individual-based simulation, a lattice-based stepping-stone simulation, and analytical theory to explore the consequences of simple topographic features on the evolution of populations undergoing a range expansion. Any bump on an otherwise flat surface has, for the most part, negative Gaussian curvature on its outside and positive Gaussian curvature on the inside, allowing us to generalise from our findings for a conical bump. We expect the temporarily increased relative genetic diversity due to local inflation, followed by decreased genetic diversity due to the formation of a kink, to be generic features for range expansions traversing a bump. 
Also expected to be seen generally is the  lensing effect that enables individuals located at either side of a bump to have an increased expected number of descendants at the population front at later times. Under the typical assumption that more distant pairs of individuals have more genetic differences, the bump's overall effect as a converging lens serves to bring into contact more genetically distinct portions of the population, more quickly. 


We have made much use of shortest-path geometrical arguments in understanding how surface curvature affects the dynamics of the range expansion. Some conditions of validity of this approach for the case of obstacles were explored in detail in Ref.\,\cite{MoebiusPCB2015}. For our purposes here, the important condition is that the size scale of the topographic features is large compared to the thickness of the front so that the front may be approximated as curvilinear expanding with constant speed. Note that geometrical optics on curved surfaces in the literal sense has been studied for a long time \cite{RighiniAppliedOptics1973} and is still a topic of active research \cite{HorsleyScientificReports2014}. Even though the analogy to geometrical optics has its limits (populations cannot pass through each other, for example), exploiting this connection to the field of optics and beyond may benefit the field of range expansions.




It is important to note that all of the effects of curved surfaces considered in this work are unchanged if the surface is inverted. An isolated bump or mountain is the same as an isolated pit or valley, from the range expansion's point of view. This fact will be important when considering topographies of greater complexity. 

Given the striking similarities between the effects of topographic bumps and nutrient-depleted obstacles, it is natural to wonder whether propagation over an arbitrary surface can be mapped onto propagation in a plane with spatially varying propagation speed -- variations created, for example, by heterogeneous nutrient availability. As explained in the Appendix, to achieve the same (projected) propagation on a planar surface as on a curved surface, the locally defined propagation speed would have to be not only heterogeneous but also anisotropic, a situation unlikely to arise naturally. Qualitatively, however, the effects of obstacles and bumps in an otherwise flat environment are very similar. Specifically, the fate of individuals depends on the \textit{a priori} random location at the front and the resulting effects can be understood as a kind of `geometry-enhanced genetic drift'.

In our focus on insights gained from geometrical optics for ensemble-averaged behaviours, we have not emphasized transverse fluctuations of sector boundaries and lineages. Fluctuations are, however, a key reason for the loss of genetic diversity. Due to coupling with the roughness of the front, fluctuations can be \textit{superdiffusive} \cite{HallatschekPNAS2007,KorolevRevModPhys2010,gralka2016}. Our findings raise the question of how to quantify the effects of fluctuations leading sector boundaries or lineages off of their initial geodesics -- possibly into the caustic, and possibly safely away from the caustic. On the plane, the overall effect of superdiffusive fluctuations is to decrease genetic diversity faster than in the case of diffusive fluctuations \cite{HallatschekPNAS2007,KorolevRevModPhys2010}. How this effect is altered by surface curvature remains to be understood.  We have shown that geometrical optics is a powerful tool for understanding the main effects of simple surface geometries, and illuminates how topographic lensing produces geometry-enhanced genetic drift. To understand range expansions in general complex environments, it will be necessary to combine this geometrical optics viewpoint with a quantitative description of fluctuations, and to develop effective descriptions for roughness at small length scales, topics we aim to address in future research.



\section{Acknowledgements}

Work by F.~Tesser was partially supported by the Nederlandse Organisatie voor Wetenschappelijk Onderzoek I (NWO-I), the Netherlands. D.A.B. thanks the American Physical Society and Sociedade Brasileira de F\'isica for supporting this research through the Brazil-U.S. Physics Ph.D. Student \& Postdoc Visitation Program. R.A.M.\ acknowledges support from FAPESP grant 2013/09357-9. The stepping-stone numerical simulations in this work were supported by the Center for Computation and Visualization, Brown University. The authors thank Sherry Chu for helpful discussions.

\clearpage

\pagestyle{plain}
\onecolumn

\begin{center} 
\textbf{\Large Appendix}  \\[5pt] 
\textbf{\large Evolution of populations expanding on curved surfaces} \\[5pt] 
{Daniel A.\ Beller, Kim M. J. Alards, Francesca Tesser,\\Ricardo A.\ Mosna, Federico Toschi, Wolfram M\"obius}
\end{center}

\newcounter{mysecnum}
\newcounter{mysubsecnum}

\newcommand{\mysection}[1]{{\vspace{11pt} \stepcounter{mysecnum}\setcounter{mysubsecnum}{1} \noindent \large \textbf{\arabic{mysecnum} \;#1\vspace{2pt}}}}
\newcommand{\mysubsection}[1]{{\vspace{11pt} \noindent \textbf{\arabic{mysecnum}.\arabic{mysubsecnum} \; #1\stepcounter{mysubsecnum} \vspace{2pt}}}}
\setcounter{tocdepth}{2}
\setcounter{equation}{0}

\section{Details of simulations}

In both individual-based and stepping-stone simulation schemes, we implemented periodic boundary conditions in the $x$-direction orthogonal to front propagation, and open boundary conditions in the $y$-direction parallel to front propagation.

Comparison to predictions of geometrical optics is based on the distance the front has propagated close to the system's boundaries in $y$-direction (averaging over a region close to the boundary to address fluctuations in front shape). In the case of individual-based simulations, the simulation was run for a given simulation time and the average front position was determined. In the case of stepping-stone simulations, the ensemble used for ensemble averages consisted of simulations within which the front has propagated to a given $y$-value close to the system's boundaries in the $x$-direction (and thus farthest from the cone).

\subsection{Individual-based simulations}

Simulation parameters were set to $\delta=1$, $D=1$, $\mu=1$, and $\lambda=1$ 
and system width to 1000. The front is defined locally as the particle most advanced in the $y$-direction in a given stretch of width $\delta$ parallel to the $y$-axis.

In Fig. 1, a Gaussian-shaped bump centered  at $(x,y)=(500,500)$ is represented by the height function $f(x,y)=300\cdot\exp\left(-((x-500)^2+(y-500)^2)/2/100^2\right)$. We additionally introduced a cutoff at a radius of $500$ around the centre beyond which $f(x,y)=0$. In Fig. 2, the surface is a cone with radius $R=300$ and height $H=300$. To compute ensemble-averaged lineages we averaged over 64 instances of the simulation.

Figs. 1B,C,D as well as Videos 1,2 were created using tachyon, the ray tracer in the package VMD \cite{VMD}. Note that in Fig. 1D and Video 2 past birth events are shown on top of the population at a given time, thus combining past events with the system's current configuration to guide the eye.

\subsection{Stepping-stone simulations}

The front is defined as the set of occupied sites with at least one free neighbouring site. For computing ensemble-averaged lineages, when the front contains more than one point at the same $x$-value, one of these points is chosen randomly. 
For each topography, 1000 simulations were run.

In Figs.\ 2D, 3A, 3C, and 4: In units of the lattice spacing, the cone has radius $R=150$, with centre at position $(x,y)=(500,400)$, and the system width is 1000. Cone height is indicated in the caption or on the axis label. In Fig 4, the calculation of $N_d(\mathrm{x})$ is made after the front has travelled a distance 1000. In Fig.\ 5, the system width is 1200, and all six cones have $H=R=125$.

\section{Computing ensemble-averaged lineages}

For computing ensemble-averaged lineage trajectories (`mean lineages') it is necessary to take distances on the curved surface into account. This requires generalising the arithmetic mean. Given a set of trajectories ${\vec{x}_i(t)}$ where $i=1\ldots N$ denotes the instance of the simulation, we define the ensemble-averaged trajectory $\vec{x}_m(t)$ as
\begin{equation}
\vec{x}_m(t) = \min_{\vec{x}} \left\{ \sum_i d\left(\vec{x},\vec{x}_i(t)\right)^2 \right\}
\label{eq:defxm}
\end{equation}
for all $t$, where $d(\vec{y},\vec{z})$ is the shortest distance from $\vec{y}$ to $\vec{z}$ on the surface as explained below. We note in passing that this is the same as calculating a generalised centre of mass of a distribution of equal-mass particles living in a curved space in the context of gravitational physics (see, e.g., Ref.~\cite{HarteInBook}). As one can easily see, this simplifies to $\vec x_m(t)=\sum_i \vec{x}_i(t)/N$ for flat surfaces as expected.

\begin{figure}
\centering
\includegraphics[width=\linewidth]{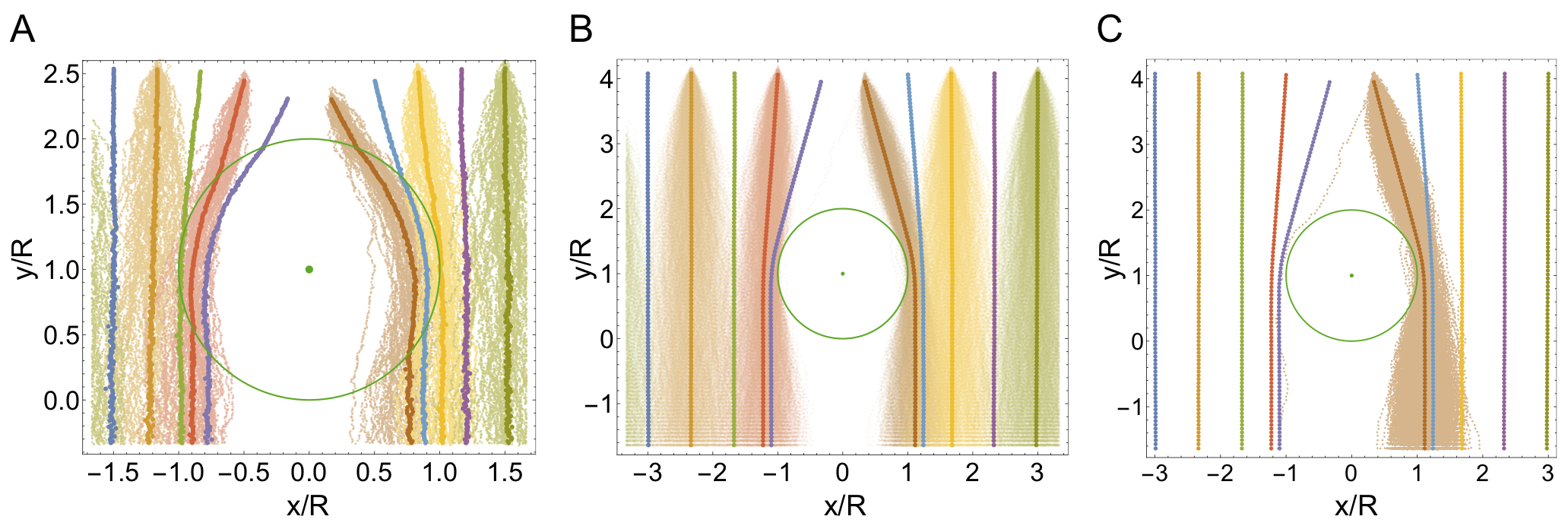}
\caption{Individual lineages (lighter-shaded points) constituting some of the ensemble-averaged lineages (heavier-shaded points) in Fig. 2C,D of the main text. Colours indicate chosen starting $x$-value of the lineages at the front. (A)~Individual-based simulations. (B) Stepping-stone simulations. (C) Same as (B) but viewing only one starting $x$-value of lineages at the front.}
\label{appendixfig1}
\end{figure}

Going back in time, individual lineages will pass over different parts of the cone or the flat surface and therefore experience different curvatures, see Fig.~\ref{appendixfig1} in this appendix. For example, in Fig.~\ref{appendixfig1}C, one lineage (shaded brown) traverses from the right to the left side of the cone, and very nearly follows the ensemble-averaged lineage from a quite different starting $x$-value. Therefore, this ensemble average may not represent a likely path for any one lineage. We only expect the ensemble-averaged lineage to represent a typical lineage trajectory minus fluctuations when the lineages remain sufficiently close. We nevertheless plotted all ensemble-averaged lineages back to the origin.

Each individual lineage consists of a number of birth events and their corresponding time and space coordinates. To obtain synchronous trajectories $\vec{x}_i(t)$, we binned the birth events in time (250 bins for individual-based simulations and 100 bins for stepping-stone simulations, respectively) and thereby chose the birth event closest to the centre of the bin for each trajectory if at least one birth event occurred in a time interval. Starting from the arithmetic mean in the projected plane or from $\vec{x}_m$ at an earlier time, we minimised the sum of squared distances (Eq.~\ref{eq:defxm}) using the SciPy implementation of the Nelder-Mead algorithm \cite{NRBook,scipy}.

\section{Computing distance between two points on surface}

\subsection{General considerations}

To compute the shortest distance $d(\vec{y},\vec{z})$ between two points $\vec{y}$ and $\vec{z}$ on a surface consisting of a plane with a conical bump, we had to distinguish several cases. (i) The case of $\vec{y}$ and $\vec{z}$ being both located on the plane is the simplest if the connecting line on the plane does not intersect with the cone: $d(\vec{y},\vec{z})=|\vec{y}-\vec{z}|$. (ii) If both $\vec{y}$ and $\vec{z}$ are located on the cone, the distance can be calculated analytically using a metric for the conical surface, as described in Ref.~\cite{MosnaPRE2012}. (iii) If (without loss of generality) $\vec{y}$ is located on the cone and $\vec{z}$ on the plane, the shortest path crosses the cone's edge at some point $\vec{v}$. To find the length of the shortest path, one needs to find the minimum of the path lengths over the full range of possible points $\vec{v}$ on the cone edge, on either side of which the geodesic is a straight line, either on the plane or on the cut-disk view of the cone. It is necessary to restrict $\vec{v}$ to the portion of the cone edge for which a straight line from $\vec{v}$ to $\vec{z}$ does not intersect the cone edge anywhere expect at the endpoint $\vec{v}$. (iv) If $\vec{y}$ and $\vec{z}$ are both located on the plane but their connecting line on the plane \textit{does} intersect with the cone, then the shortest path between $\vec{y}$ and $\vec{z}$ crosses the conical portion of the surface. Finding the shortest path requires finding two points $\vec{v}_1$, $\vec{v}_2$ where the geodesic intersects the cone edge. In principle, $\vec{v}_1$ and $\vec{v_2}$ are determined by two rules discussed in Ref.~\cite{MosnaPRE2012}: A version of Snell's law requires that at each of these two points the geodesic makes the same angle with the cone edge's tangent both inside and outside the cone (as measured on the cone and on the plane, respectively); and Clairaut's relation for axisymmetric surfaces requires that this angle be the same at both $\vec{v}_1$ and $\vec{v}_2$. These constraints together determine both $\vec{v}_1$ and $\vec{v}_2$ implicitly, and one can in principal solve for their values numerically. (A similar procedure could also be applied in case (iii).) As in case (iii), $\vec{v}_1$ and $\vec{v}_2$ are restricted to the portions of the cone edge for which the lines connecting $\vec{v_1}$ to $\vec{y}$ and $\vec{v_2}$ to $\vec{z}$ intersect the cone edge only at the endpoints $\vec{v}_1$ and $\vec{v}_2$, respectively. 

\subsection{Numerical implementation}

As indicated above, the distance between two points located in the simulation domain can be computed either analytically (cases (i) and (ii)) or through solving an implicit equation (cases (iii) and (iv)). Instead of numerically solving the implicit equation(s), we minimised the total path length over all possible points $\vec{v}$ or $\vec{v}_1$ and $\vec{v}_2$ in the allowed portions of the cone edge discussed above. 
Since the cone edge is a circle, the sought-after distance-minimizing points on the cone edge can be parametrised by their polar angle. Finding the angle(s) that minimise the distance between two points thus represents a one-dimensional minimisation problem for case (iii) and a minimisation in two dimensions for case (iv).

We searched for the minimum distance using the SciPy implementation of the Nelder-Mead algorithm \cite{NRBook,scipy} starting from very close to the boundaries of the interval of allowed angle(s) (2 starting conditions for case (iii), 4 starting conditions for case (iv)) and making use of a heuristic penalty term to constrain angles to their allowed range. The true distance was taken to be the smallest of the various local minima found using these various starting points in the search space. To test that the global minimum has not gone undetected, we also used a grid of 100 points of $\vec{v}$ (case (iii)) or 10000 pairs of $(\vec{v}_1,\vec{v}_2)$ (case (iv)), computed the resulting distance, and compared with the minimum found as described above.

\subsection{Treatment of periodic boundary conditions}

To take into account the periodic boundary conditions used, one of the points $\vec{y}$ or $\vec{z}$ was translated by $-L$, $0$, or $+L$ in the $x$-direction, where $L$ is the system width. The  distance between $\vec{y}$ and $\vec{z}$ in the presence of periodic boundary conditions then is the minimum of all three cases considered. This can in principle result in complex paths involving more than one cone. Computing the distance in the projected plane as a lower limit for the distance on the surface revealed that these more complex cases did not need to be considered in any of the scenarios presented in this work.

\section{Relation to flat heterogeneous environments}

To map the dynamics on the surface to a heterogeneous flat environment, the parameters describing this environment need to be anisotropic in general. This can be seen from the perspective of the microscopic rules, the front dynamics, and the geometry of the surface. The algorithm used for the individual-based simulation computes particle motion in the projected plane and this way provides the desired mapping. The isotropic diffusion in the tangential plane relates to anisotropic diffusion in the projected plane. Similarly, from the perspective of the front, isotropic propagation of a radial front in a tilted plane maps to anisotropic propagation in the projected plane. Last but not least, a fundamental concept from differential geometry is that a surface's radius of curvature at a given point generally depends on the choice of direction in the local tangent plane \cite{DoCarmo2016Book}; quantities such as the mean and Gaussian curvatures are ways of averaging over these directions.

\end{document}